\documentclass[referee]{aa} 
\usepackage{graphicx}
\usepackage{natbib}

\bibpunct[]{(}{)}{;}{a}{}{,}
\def\ltsima{$\; \buildrel < \over \sim \;$}
\def\simlt{\lower.5ex\hbox{\ltsima}}
\def\gtsima{$\; \buildrel > \over \sim \;$}
\def\simgt{\lower.5ex\hbox{\gtsima}}

\def\cm2{\mbox{$\mbox{cm}^{-2}$}}
\def\cm3{\mbox{$\mbox{cm}^{-3}$}}
\def\h2{\mbox{$_{\mbox{\tiny H2}}$}}
\def\figs_these_arnaud{figs_these_arnaud}

\begin{document}
   \title{First 450 $\mu$m dust continuum mapping of the massive star-forming region NGC~3576 with the 
   P-ArT\'eMiS bolometer camera\footnote{This publication is based on data acquired with the Atacama 
Pathfinder Experiment (APEX) in ESO program 080.C-0722. 
APEX is a collaboration between the Max-Planck-Institut f\"ur Radioastronomie, the European
Southern Observatory, and the Onsala Space Observatory.}~$^,$
\footnote{Tables~\ref{ngc3576_tab1} \& ~\ref{ngc3576_tab2} and Figs.~\ref{simba_map} \&~\ref{ratio_map} 
are only available in electronic form.} 
}

   \author{Ph.~Andr\'e\inst{1}
          \and
          V.~Minier\inst{1}
           \and
          P.~Gallais\inst{1}
          \and
          V.~Rev\'eret\inst{2,1}          
          \and
	 J.~Le~Pennec\inst{1}
          \and
          L.~Rodriguez\inst{1}
          \and
          O.~Boulade\inst{1}
          \and
          E.~Doumayrou\inst{1}
          \and
          D.~Dubreuil\inst{1}
          \and
          M.~Lortholary\inst{1}
          \and
          J.~Martignac\inst{1}
	\and
          M.~Talvard\inst{1}          
          \and 
          C.~De~Breuck\inst{3}
	\and
          G.~Hamon\inst{1}          
	 \and 
          N.~Schneider\inst{1}
          \and 
          S.~Bontemps\inst{1}
          \and 
          P.O.~Lagage\inst{1}
          \and 
          E.~Pantin\inst{1}
          \and
          H.~Roussel\inst{4}
          \and 
          M.~Miller\inst{5}
	\and
          C.R.~Purcell\inst{6}
 	\and
          T.~Hill\inst{7}                   
          \and 
          J.~Stutzki\inst{5}           
          }

   \offprints{pandre@cea.fr}

   \institute{Laboratoire AIM, CEA/DSM--CNRS--Universit\'e Paris Diderot, IRFU/Service d'Astrophysique, C.E. Saclay,
              Orme des Merisiers, 91191 Gif-sur-Yvette, France
	 \and
             European Southern Observatory,  Casilla 19001, Santiago 19, Chile
 	 \and
	   European Southern Observatory,  Karl Schwarzschild Str. 2, 85748 Garching bei Munchen, Germany
 	 \and
	   Institut d'Astrophysique de Paris, Universit\'e Pierre \& Marie Curie,   
	   98b Bd Arago, 75014 Paris, France
	 \and
	   I. Physik. Institut, Universit\"at zu K\"oln, Z\"ulpicher Str. 77, 50937 K\"oln, Germany
 	 \and
	   University of Manchester, Jodrell Bank Observatory, Manchester M13 9PL, UK
	 \and 
	   School of Physics, University of Exeter,  Stocker Rd, EX4 4QL, Exeter, UK 
             }

   \date{Received ; accepted }

   \abstract
  {The earliest phases of massive star formation are currently much debated. } 
  {In an effort to make progress, we took a census of Class~0-like protostellar dense cores in the NGC~3576 region, 
  one of the nearest and most luminous embedded sites of high-mass star formation in the Galaxy.}
  {We used the P-ArT\'eMiS bolometer camera on 
  the APEX telescope to produce the first 450~$\mu$m dust continuum map of the  filamentary dense clump associated with NGC~3576. 
  }
  {Combining our 450~$\mu$m observations with existing data at other wavelengths, we have identified seven
 massive protostellar sources  along the NGC~3576 filament and placed 
  them in the $M_{env}$--$ L_{bol}$ evolutionary diagram for protostars.  
   }
  {Comparison with theoretical evolutionary tracks suggests that these seven protostellar sources 
  will evolve into massive stars with masses $M_\star \sim $~15--50$\, M_\odot $.
  Four sources are classified as candidate high-mass Class~0 objects, two sources  as massive 
  Class~I objects, and one source appears to be at an intermediate stage.}

   \titlerunning{P-ArT\'eMiS 450 $\mu$m mapping of NGC~3576}
   \authorrunning{Andr\'e et al.}
   \maketitle
%

\section{Introduction}
\label{intro}

The process of massive star formation remains poorly understood
and is the subject of major ongoing debate (Zinnecker \& Yorke 2007). 
From the observational perspective, there is a growing body of evidence 
to suggest 
that most, if not all, high-mass ($M_\star > 8\, M_\odot $) stars form in a 
clustered environment (
de Wit et al. 2005). There is no widely 
accepted evolutionary scenario for high-mass star formation, although a tentative sequence of empirical phases 
has been identified: from infrared dark cloud cores, to hot cores, hypercompact to ultracompact,  
compact and classical HII regions. 
On the theoretical front, it is unclear whether high-mass stars form by direct cloud core collapse and subsequent 
accretion as a result of a scaled-up version of the process believed to be at work in low-mass protostars  
(McKee \& Tan 2003), or by other processes such as competitive accretion and coalescence of 
intermediate-mass protostars in dense protoclusters (Bonnell et al. 2004).

Decisive progress in this field will require high-resolution kinematical and modeling studies of high-mass protostellar 
objects and their embedded, cluster-forming environments at (sub)millimeter wavelengths. While such comprehensive studies
are becoming feasible in the intermediate-mass regime  (cf. Peretto et al. 2006, 2007), they are virtually impossible 
in the high-mass regime due to the lack of angular resolution and mapping speed of current submillimeter facilities. 
A major step forward will be provided by the upcoming ALMA interferometer when it becomes fully operational around 2013.
In order to prepare for future ALMA studies, it is important to identify a number of candidate massive 
protostars and embedded protoclusters in the southern hemisphere. 
While several SIMBA 1.2~mm continuum surveys  are available (e.g. Hill et al. 2005, 
Beltr\'an et al. 2006), current searches for massive protostars in the southern sky lack both the angular resolution 
and submillimeter spectral coverage   
to provide an adequate characterization of the detected sources.

At a distance of 2.4~kpc (Persi et al. 1994), NGC~3576 (also known as RCW 57 and G291.27-0.70) is one of the most luminous  
giant HII regions in our Galaxy at optical, infrared, and radio wavelengths (e.g. De Pree et al. 1999). 
It is embedded at the center of a filamentary dense clump observed in both 1.2~mm dust continuum emission 
(Hill et al. 2005) and tracers of dense molecular gas such as NH$_3$ (Purcell et al. 2008).   
A deeply embedded near-IR cluster of more than 130 young stellar objects (YSOs) is known to be associated 
with this region (Persi et al. 1994). In addition, two 6.7~GHz methanol masers 
and a total of nine 
22~GHz water masers (Caswell 2004, Purcell et al. 2008) have been detected along the extent of the cluster-forming clump.
The NGC~3576 region is therefore an excellent laboratory for studying the embedded phases of massive star 
formation. Due to its southern declination, NGC~3576 is a potential target of choice for future high-resolution 
investigations with ALMA.

Here, we report the results of  the first 450~$\mu$m dust continuum mapping observations of the NGC~3576 star-forming 
region,  taken with the P-ArT\'eMiS bolometer camera on the APEX 12-m telescope. 
Section~\ref{obs_set} describes the instrument and provides details about the observing run and data reduction. 
Section~\ref{obs_ana} presents our mapping results  which are discussed in Section~\ref{dis}.

\section{Observations and data reduction}
\label{obs_set}

Our 450~$\mu$m observations of NGC~3576 were made in March 2007 and November 2007 with the P-ArT\'eMiS bolometer 
array on the Atacama Pathfinder Experiment (APEX) telescope located at an altitude of 5100~m at Llano de Chajnantor in Chile. 
P-ArT\'eMiS is a $16 \times 16$--pixel prototype of the ArT\'eMiS bolometer camera currently being built by CEA Saclay for APEX. 
ArT\'eMiS\footnote{See http://www.apex-telescope.org/instruments/pi/artemis/\\
ArT\'eMiS stands for ``ARchitectures de bolom\`etres pour des TElescopes \`a grand champ de vue dans le domaine sub-MIllim\'etrique au Sol'' in French.}
is a large-format camera with a total of 5760 bolometric pixels which will operate simultaneously at 450~$\mu$m,  350~$\mu$m,  
and 200~$\mu$m in the Cassegrain cabin of APEX (e.g. Talvard et al. 2006, 2008).
It uses the technology successfully developed by CEA for the PACS photometer instrument in the 60--210~$\mu$m wavelength regime 
on the $Herschel$ Space Observatory (e.g. Simoens et al. 2004, Billot et al. 2006). Unlike classical bolometer arrays, such as LABOCA 
on APEX, the PACS and (P-)ArT\'eMiS instruments do not use feedhorns to concentrate the incoming submillimeter radiation, but act 
more like CCD cameras in the optical domain.
The advantage of such a filled-array design is that the entire field of 
view of the camera is instantaneously fully sampled, and thus ``jiggling'' is not required to produce fully-sampled images.
The P-ArT\'eMiS prototype was built with the objective of demonstrating the feasibility and potential of the full ArT\'eMiS instrument. 
P-ArT\'eMiS was first tested at 450~$\mu$m on the KOSMA 3m telescope 
in March 2006 and January 2007. It was then taken to Chile for initial tests on APEX during the second half of March 2007
under technical time allocated by the APEX Board, and then during ESO and Onsala scientific time in November 2007. 
P-ArT\'eMiS uses a detector array produced as part of the $Herschel$-PACS project, but not selected for the space mission, 
which was modified to operate at 450~$\mu$m by adding a silicon-based filtering system on its surface (cf. Rev\'eret et al. 2006). 
The P-ArT\'eMiS bolometers are operated at 300~mK inside a conventional He-refilling cryostat. 
During the 2007 runs, P-ArT\'eMiS was installed by our team at the visitor focus in the Nasmyth ``A'' cabin of the APEX telescope.

Eight individual maps, corresponding to a total effective integration time of 3.6~hr,  were obtained with P-ArT\'eMiS at $450\, \mu$m toward 
the NGC~3576 region using a total-power, on-the-fly scanning mode. Each of these maps consisted of a series of scans in either Azimuth or Right Ascension 
taken with a scanning speed of 16\arcsec /sec or 20\arcsec /sec. The cross-scan step between consecutive scans was either 
2\arcsec ~or 4\arcsec . The sizes of the maps ranged from $4'  \times 4'$ to $6'  \times 5'$. The atmospheric opacity at zenith 
was monitored by taking skydips with P-ArT\'eMiS and was found to be between 0.6 and 0.8 at $\lambda = 450\, \mu$m, 
corresponding to an amount of precipitable water vapor (PWV) between $\sim $~0.5~mm and $\sim $~0.7~mm. No dedicated pointing 
model was derived for P-ArT\'eMiS, but the pointing corrections proved to be very similar to, i.e. within $\sim 10\arcsec $ 
of, the corrections appropriate for the APEX-2A instrument. Pointing, focus, and calibration measurements were achieved 
by taking both short `spiral' scans and longer on-the-fly beam maps of Mars and Saturn. 
We estimated the absolute pointing accuracy to be $\sim 5\arcsec $ and the absolute calibration uncertainty to be $\sim 30\% $.  
The noise equivalent flux density (NEFD) per detector was  
$\sim $~2~Jy.s$^{1/2}$ .
and the pixel separation between detectors on the sky was $\sim 6\arcsec $. The individual maps of NGC~3576 
were registered with a relative positional accuracy of $\sim 2\arcsec $ and a relative calibration 
uncertainty of $\sim 10\% $. 
As estimated from our maps of Mars (whose angular diameter was $\sim 14\arcsec $ in November 2007), 
the main beam had a full width at half maximum (FWHM) $\sim 10\arcsec $ and contained $\sim 60\% $ of the power, the 
rest being distributed in an ``error beam'' extending up to an angular radius of $\sim 80\arcsec $. 
On-line data reduction was performed with the BoA software 
developed for LABOCA (Schuller, Beelen et al., in prep.). 
Off-line data reduction, including baseline subtraction, removal of correlated skynoise and $1/f$ noise, as well as 
subtraction of uncorrelated $1/f$ noise using a method exploiting the high level of redundancy in the data 
(Roussel et al., in prep.), was performed with in-house IDL routines. 

\section{Mapping results and analysis}
\label{obs_ana}

\begin{figure} [ht]
\centerline{\resizebox{1.0\hsize}{!}{\includegraphics[angle=270]{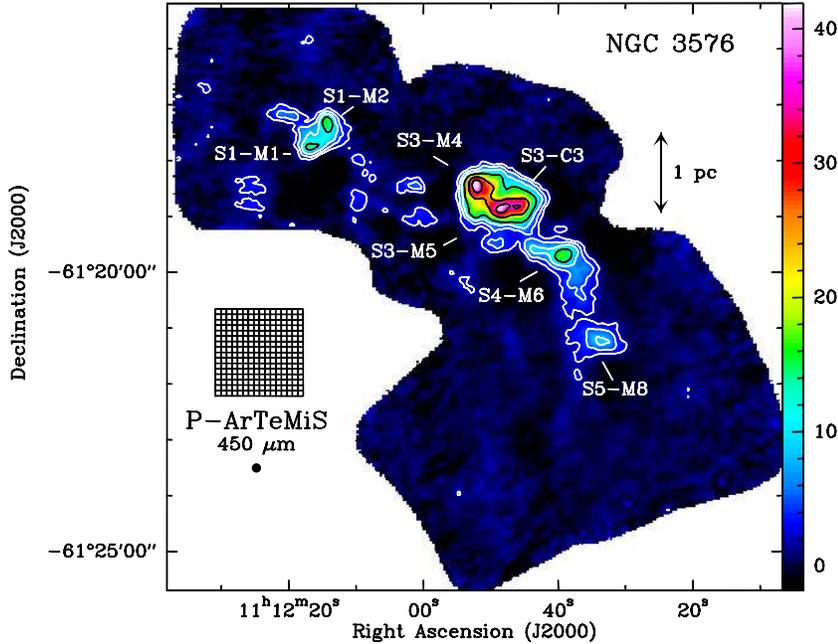}}}
\caption[P-ArT\'eMiS map]
{Total-power 450~$\mu$m dust continuum mosaic of the massive star-forming region NGC~3576 taken with P-ArT\'eMiS on APEX.
The rms in this mosaic is $\sim 0.45 $ Jy/10\arcsec -beam. The peak is $\sim 47 $ Jy/10\arcsec -beam. 
Contour levels are 2, 4.5, 8, 13, 25, and 37.5~Jy/10\arcsec -beam. The color scale shown on the right is in 
Jy/10\arcsec -beam.
The grid shown on the left corresponds to the instantaneous field of view of the $16\times16$~-pixel P-ArT\'eMiS array. 
The 10\arcsec ~(FWHM) beam size is also indicated. }
\label{partemis_map}
\end{figure}

Co-adding the eight individual P-ArT\'eMiS maps of NGC~3576, we obtained the $450\, \mu$m mosaic shown in 
Fig.~\ref{partemis_map}. At least seven localized dust emission peaks can be identified in this mosaic, which have been 
labeled here following the nomenclature of Purcell et al. (2008). 
They are distributed in four $\sim $~parsec-scale clumps labeled S1, S3, S4, and S5. 
The positions and flux densities measured for these compact P-ArT\'eMiS sources are given in online Table~\ref{ngc3576_tab1}. 

While the $450\, \mu$m emission 
peaks are well correlated with the 1.2~mm peaks detected by Hill et al. (2005) with SIMBA on the SEST, 
the factor of $\sim 2$ gain in angular resolution provided by P-ArT\'eMiS 
and the absence of confusion by free-free emission (see below) 
allowed us to resolve at least two of the SIMBA peaks (S1 and S3) into multiple $450\, \mu$m dust continuum sources 
(see online Fig.~\ref{simba_map}). 

To further compare our P-ArT\'eMiS $450\, \mu$m mosaic with the SIMBA 1.2~mm map obtained by Hill et al. (2005), we 
smoothed the P-ArT\'eMiS data to the 22\arcsec ~resolution of the SIMBA data, regridded the maps onto the same grid, and 
produced a map of the 450~$\mu$m to 1.2~mm flux density ratio over the region where significant 1.2~mm continuum 
emission was detected by SIMBA. The resulting ratio map is shown in online Fig.~\ref{ratio_map}. 
Over most of the area with significant emission at both $450\, \mu$m and 1.2~mm, the 
$S_{450\mu {\rm m}}^{22\arcsec}/S_{1.2{\rm mm}}^{22\arcsec}$ ratio ranges between $\sim 20$ and $\sim 40$.
These values are consistent with the $S_\nu \propto B_\nu (T_d)\, \nu^\beta $ spectral dependence expected for optically thin emission 
from dust  with an emissivity index $\beta = 2$ 
(e.g. Hildebrand 1983) and a temperature $T_d \sim $~10--30~K. 
In the eastern part of the S3 clump, however, the observed $S_{450\mu {\rm m}}^{22\arcsec}/S_{1.2{\rm mm}}^{22\arcsec}$ 
ratio is much lower, with values as small as $\sim $~1.5--7.5 (cf. 
online 
Fig.~\ref{ratio_map}). Such values of the flux ratio are inconsistent 
with dust emission unless the dust temperature is unrealistically low, $T_d \simlt $~5--7~K,  for a parsec-scale clump. 
This reflects the fact that the 1.2~mm emission observed with SIMBA toward the S3 clump is strongly 
contaminated by free-free radiation from the RCW~57 HII region (marked by white contours in Fig.~\ref{ratio_map}). 
Extrapolating the 1.3~cm continuum emission measured by ATCA (Purcell et al. 2008) to 1.2~mm with the 
$S_\nu \propto \nu^{-0.1}$  spectrum of optically thin free-free radiation, we estimate that free-free emission 
contributes more than 50\% of the 1.2~mm flux densities observed with SIMBA in the eastern part of the S3 clump.

\onlfig{2}{
\begin{figure*} [!htbp]
\centerline{\resizebox{0.9\hsize}{!}{\includegraphics[angle=270]{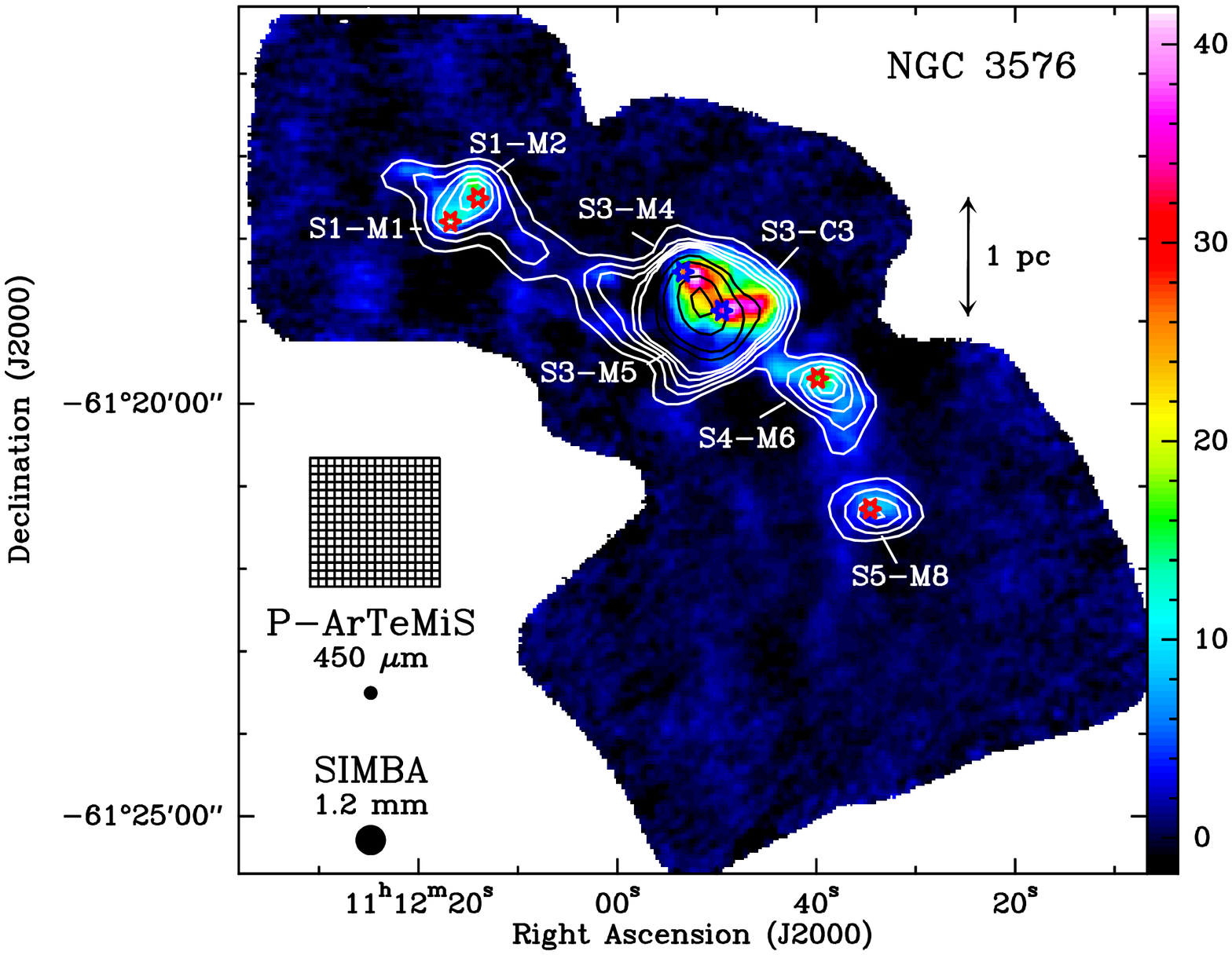}}}
\caption[SIMBA/P-ArT\'eMiS map]
{Comparison with the SIMBA/SEST 1.2mm continuum map of NGC~3576 
(Hill et al. 2005), overlaid as contours on the P-ArT\'eMiS 450~$\mu$m color image.
The white SIMBA contours go from 0.3 to 1.5 by 0.3 Jy/22\arcsec -beam, and the black SIMBA contours are  
3, 4.5, 9, and 13.5 Jy/22\arcsec -beam. Blue stars mark the positions of two 6.7~GHz methanol maser sites 
(Caswell 2004), while red stars mark 22~GHz water maser sites (Purcell et al. 2008). 
Note that at least two of the SIMBA peaks (S1 and S3) are resolved into multiple P-ArT\'eMiS sources.}
\label{simba_map}
\end{figure*}
} 

\onlfig{3}{
\begin{figure*} [!htbp]
\centerline{\resizebox{0.8\hsize}{!}{\includegraphics[angle=270]{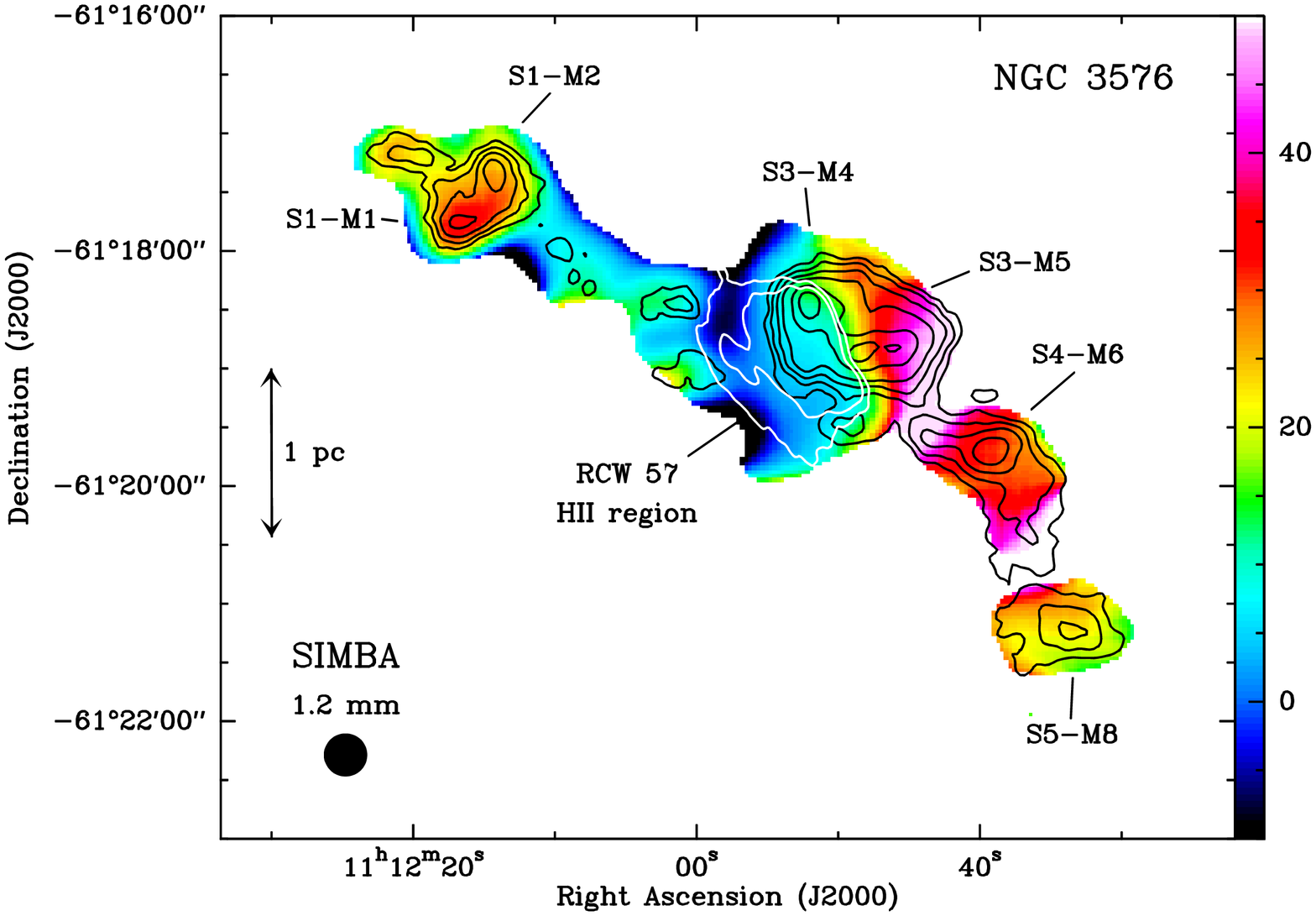}}}
\caption[P-ArT\'eMiS/SIMBA ratio map]
{Color-scale image showing the variation of the 450~$\mu$m to 1.2~mm flux density ratio across the NGC~3576 complex. 
The range of observed intensity ratios is indicated by the color scale on the right. 
The (FWHM) angular resolution of this image is the same as that of the SIMBA map (Hill et al. 2005), i.e., $\sim 22\arcsec $.
The P-ArT\'eMiS 450-$\mu$m  map shown in Fig.~\ref{partemis_map} is superimposed as black contours. The white 
contours outline the extent of the RCW~57 HII region as imaged at 1.3~cm with ATCA (Purcell et al. 2008).}
\label{ratio_map}
\end{figure*} 
}

%
\subsection{Detection of embedded protostars and derived properties}
\label{det}

Two of the seven compact P-ArT\'eMiS sources listed in 
online 
Table~\ref{ngc3576_tab1}, S3-M4 and S3-M5, are closely 
associated with 6.7~GHz methanol masers (see Fig.~\ref{simba_map}) and bright MSX sources. 
The four P-ArT\'eMiS sources S1-M1, S1-M2, S4-M6, S5-M8 were not detected by MSX but are closely associated with 
22~GHz water masers (see Fig.~\ref{simba_map}). The seventh P-ArT\'eMiS source, 
labeled S3-C3 here, also remained undetected by MSX but coincides with a warm ($T_{NH_3} \sim $~30--40~K) ammonia core 
recently identified by Purcell et al. (2008) with ATCA (and labeled C3 by these authors). These associations indicate that the 
$\sim $~0.1--0.4~pc dusty dense cores detected with P-ArT\'eMiS contain at least one protostellar object each. 


While the $\sim 10\arcsec $ ~(HPBW) angular resolution of P-ArT\'eMiS at 450~$\mu$m
is hardly sufficient to resolve individual condensations/envelopes within the dust cores 
at the distance of NGC~3576, 
our data can be used to estimate the masses and bolometric luminosities of the dominant 
protostellar sources as follows:

\onltab{1}{
\begin{table*} 
\centering 
\caption{Protostellar sources detected with P-ArT\'eMiS in NGC~3576} 
\label{ngc3576_tab1} 
\begin{tabular}{lccccccl} 
\hline\hline 
Source & \multicolumn{2}{c}{Coordinates}  & $S_{450\mu {\rm m}}^{peak} $ & $S_{450\mu {\rm m}}^{30\arcsec} $ &  $\frac{S_{450\mu {\rm m}}}{S_{1.2{\rm mm}}} $ & $T_d$  & Remarks \\
             & $\alpha_{2000}$ & $\delta_{2000}$ & {(Jy/beam)} & { (Jy)} & & { (K)} &  \\  
             & 11$^{\mbox{h}}$ & $-61^\circ$ &  &  &   &   &    \\ 
(1) & (2) & (3) & (4) & (5) & (6) & (7) & (8) \\  
\hline 
S1-M1    & 12$^{\rm{m}}$16.6$^{\rm{s}}$  & 17\arcmin 51\arcsec  &  $15. \pm 0.9$   &  $45Ê\pm 2.5$   & $34Ê\pm 2$ &    $21 \pm 3$    &   H$_2$O maser, no MSX  \\ 

S1-M2    & 12$^{\rm{m}}$14.2$^{\rm{s}}$  & 17\arcmin 26\arcsec  &  $18. \pm 0.7$   &  $53 \pm 2$    &  $27 \pm 2$ &   $16 \pm 2$    &   H$_2$O maser, no MSX  \\ 
\hline
S3-M4    & 11$^{\rm{m}}$53.2$^{\rm{s}}$  & 18\arcmin 27\arcsec  &  $51. \pm 0.5$   & $169 \pm 1.5$    & $\simgt 20^a$  &    $\simgt 12$      &  CH$_3$OH maser, MSX  \\

S3-M5     & 11$^{\rm{m}}$49.65$^{\rm{s}}$  & 18\arcmin 52\arcsec  &  $50. \pm 0.5$   & $199 \pm 2$  & $\simgt 25^a$  &   $\simgt 15$    &   CH$_3$OH maser, MSX  \\

S3-C3    & 11$^{\rm{m}}$47.5$^{\rm{s}}$  & 18\arcmin 49\arcsec  &  $45. \pm 0.5$  & $192 \pm 2$   & $40 \pm 2$ &   $33 \pm 4$   &   NH$_3$ peak (C3), no MSX  \\
\hline
S4-M6    & 11$^{\rm{m}}$40.5$^{\rm{s}}$  & 19\arcmin 43\arcsec  &  $15.5 \pm 0.5$  &  $51 \pm 1.5$   & $31\pm 1$ &  $19 \pm 1$  &   H$_2$O maser, no MSX   \\

S5-M8    & 11$^{\rm{m}}$34.3$^{\rm{s}}$  & 21\arcmin 17\arcsec  &  $11. \pm 0.5$  &  $33 \pm 1$    & $23 \pm 2$ &   $13.5\pm 2$  &   H$_2$O maser, no MSX  \\
\hline 
\end{tabular}
\vspace*{-0.45ex}
\begin{list}{}{}
\item[$^a$]{Lower limit to $\frac{S_{450\mu {\rm m}}}{S_{1.2{\rm mm}}} $ for the dust source after correcting $S_{1.2{\rm mm}}$ for free-free emission. } 
\end{list} 
\end{table*}
}


Assuming optically thin dust continuum emission at 450~$\mu$m, the gas masses $M$ of the sources were derived from the measured 
flux densities  $S_{450\mu {\rm m}}$ using $ M = \frac{S_{450\mu {\rm m}}\, d^2} {\kappa_{450\mu {\rm m}}\, B_{450\mu {\rm m}}(T_d) }, $
where $\kappa_{450\mu {\rm m}}$ is the dust opacity per unit (gas~$+$~dust) mass column density  at $\lambda = 450\, \mu {\rm m}$, 
and $B_{450\mu {\rm m}}(T_d) $ is the Planck function $B_\nu (T_d) $ for the dust temperature $T_d$. We adopted 
$\kappa_{450\mu {\rm m}} = 0.04\ \rm{cm}^{2}\, \rm{g}^{-1} $, which is consistent with the dust opacity law avocated 
by Hildebrand (1983) and  appropriate in regions of moderately high gas densities 
($n_{H_2} \sim 10^5\, \rm{cm}^{-3}$ -- cf. Henning et al. 1995). 
The mean dust temperature of each core was estimated from the observed 450~$\mu$m to 1.2~mm flux density ratio (cf. Fig.~\ref{ratio_map}) 
assuming optically thin dust emission with an emissivity index $\beta = 2$ (Hildebrand 1983). The results, given 
in Col.~7 of Table~\ref{ngc3576_tab1}, generally agree with the gas kinetic temperatures derived from the NH$_3$ observations 
of Purcell et al. (2008). Because of free-free contamination in the eastern part of the S3 clump (see above), 
our method provides only  lower limits to the dust temperatures of S3-M4 and S3-M5. For these two cores, we thus relied on 
the gas temperature estimates of Purcell et al. (2008). The adopted dust temperatures and the corresponding core masses 
are listed in Col.~2 and Col.~3 of online Table~\ref{ngc3576_tab2}, respectively. 
Column~4 provides an estimate of the envelope mass, $M_{env}^{0.03\rm{pc}}$, within a diameter of 0.03~pc of the 
{\it dominant} protostellar object in each core (where 0.03~pc $\sim 6000$~AU is the typical fragmentation lengthscale 
observed in nearby cluster-forming regions -- cf. Peretto et al. 2007). 
$M_{env}^{0.03\rm{pc}}$ was obtained by first calculating the mass corresponding to the peak flux density in 
the 10\arcsec ~beam (given in Col.~4 of Table~\ref{ngc3576_tab1} and corresponding to a $\sim$~0.12~pc 
aperture at 2.4~kpc), and then applying a scaling factor of $0.03/0.12 = 1/4$ to that mass, 
assuming a $\rho \propto r^{-2}$ density distribution so that the measured mass scales as the aperture size.
A mean column density and a mean volume density 
for each core are also given in Col.~8 and Col.~9 of Table~\ref{ngc3576_tab2}.

The bolometric luminosity $L_{bol}$ of the {\it dominant} protostellar object in each core 
was estimated by fitting the observed SIMBA, P-ArT\'eMiS, and MSX flux densities
(or upper limits) with the large grid of protostellar spectral energy distributions (SEDs) computed by Robitaille et al. (2006, 2007), 
using their online SED fitting tool. Unfortunately, no $Spitzer$-MIPS observations of NGC~3576 are available yet, consequently 
resulting in poorly-sampled  SEDs and large uncertainties on $L_{bol}$. Coupling a SED fitting approach with our 
P-ArT\'eMiS 450~$\mu$m results nevertheless allowed us to derive a useful range of possible $L_{bol}$ values for each 
protostellar core, which is given in Col.~5 of Table~\ref{ngc3576_tab2}.


%
\onltab{2}{
\begin{table*} 
\centering 
\caption{Derived properties of the NGC~3576 protostellar sources} 
\label{ngc3576_tab2} 
\begin{tabular}{lcccccccc} 
\hline\hline 
Core & $T_d$ & $M_{core}^{0.35\rm{pc}}$ & $M_{env}^{0.03\rm{pc}}$ & $L_{bol} $ &  $M_{env}/L_{bol}^{0.6} $ & $T_{bol} $ & $<N_{H_2}>_{0.35\rm{pc}}$ &   $<n_{H_2}>_{0.35\rm{pc}}$    \\
             & (K) & ($M_\odot$) & ($M_\odot$) & ($L_\odot$) & ($M_\odot /L_\odot ^{0.6} $) & (K) & ($10^{23}\, \rm{cm}^{-2}$) &  ($10^{5}\, \rm{cm}^{-3}$)\\  
(1) & (2) & (3) & (4) & (5) & (6) & (7)  & (8) & (9) \\  
\hline 
S1-M1    &    21     &     250    & 21 & 500--$10^4$  &  0.1--0.5     &  $\simlt 80$  &  $1.4 $   & $2.0 $  \\ 

S1-M2    &    16     &     540    & 45 & 500--$10^4$  &  0.2--1.1   &  $\simlt 70$  &  $3.0$   &  $4.2 $ \\ 

\hline
S3-M4    &    35     &     400    & 30 & 0.9--3~$\times 10^5$  &  0.01--0.03   & 120--170  &  $2.2 $  &  $3.1 $ \\

S3-M5     &    35   &     460   &  29 & 0.5--2.5~$\times 10^5$  &  0.02--0.04  & 110--160  &  $2.6 $  & $3.6 $   \\

S3-C3   &    33   &    490   &  29 & 0.5--10~$\times 10^4$  &  0.03--0.2   &   $\simlt 60$   &  $2.7 $  &  $3.8 $   \\
\hline

S4-M6    &    19     &   350   & 26 & 500--$10^4$  &  0.1--0.6   &  $\simlt 80$  &  $2.0 $  &   $2.7 $  \\

S5-M8    &    13.5   &   500   & 42 & 500--$10^4$  &  0.15--1.0   &  $\simlt 90$  & $2.9 $   &  $4.0 $  \\
\hline 
\end{tabular}
\vspace*{-0.45ex}
 \begin{list}{}{}
 \item[]{Notes: Column (3) gives the core mass derived from $S_{450\mu {\rm m}}^{30\arcsec} $ integrated over a 30\arcsec ~diameter (Col.~5 of Table~\ref{ngc3576_tab1}); 
 uncertainty is a factor of $\sim 2$ on either side due to the uncertain value of $\kappa_{450\mu {\rm m}}$.
 Column (4) is the envelope mass inferred for the dominant object in each core (see text); estimated uncertainty is a factor of $\sim 4$ on either side.
 Column (5) gives the bolometric luminosity estimated from $S_{450\mu {\rm m}}^{30\arcsec} $,  $S_{1.2{\rm mm}}^{30\arcsec} $, and MSX flux densities (or upper limits),
 using the fitting routine of Robitaille et al. (2007).} 
 \end{list} 
\end{table*} 
}

\begin{figure} [!htbp]
\centerline{\resizebox{0.8\hsize}{!}{\includegraphics[angle=270]{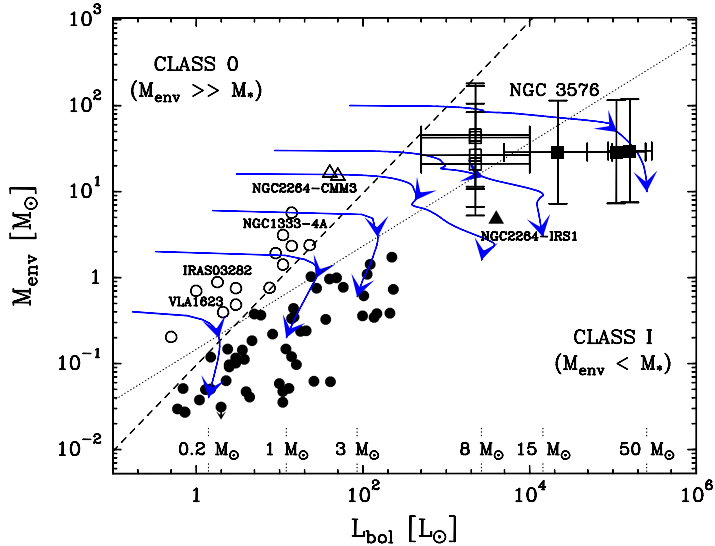}}}
\caption[menv-Lbol evolutionary diagram]
{Envelope mass versus bolometric luminosity diagram comparing the locations of the  
protostellar sources of NGC~3576 
(filled and open squares with error bars) with the positions of low-mass Class~I (filled circles)
and Class~0 objects (open circles) 
(see Andr\'e et al. 2000), as well as intermediate-mass protostars in NGC~2264 
(triangles -- Peretto et al. 2007, Maury et al. in prep.). 
Model protostellar evolutionary tracks, computed for various final stellar masses (indicated above the $L_{bol}$ axis) 
assuming the accretion/luminosity history described in the text, are superimposed. 
Evolution proceeds from the upper left to the lower right as indicated by two arrows on each track, 
plotted when 50\% and 90\% of the final stellar mass has been accreted, respectively.
The straight  lines show two $M_{env}$--$L_{bol}$ relations marking the conceptual border zone 
between the Class~0 ($M_{env} > M_\star  /\epsilon $) and the Class~I ($M_{env} < M_\star /\epsilon $) stage 
(where $\epsilon $ is the local star formation efficiency). 
The dashed line is such that $M_{env} \propto L_{bol}$ (cf. 
Andr\'e \& Montmerle 1994), 
while the dotted relation follows $M_{env} \propto L_{bol}^{0.6}$ as suggested by 
the accretion scenario adopted in the tracks.
}
\label{menv_lbol}
\end{figure}

\section{Discussion and conclusions}
\label{dis}

To put our present results in the context of similar observations of lower-mass 
Class~0 and Class~I protostellar objects and to facilitate comparison, Fig.~\ref{menv_lbol} shows the locations of the 
{\it dominant} sources of NGC~3576 in the $M_{env}$--$L_{bol}$ evolutionary diagram for protostars 
(cf. Andr\'e \& Montmerle 1994, Saraceno et al. 1996, Andr\'e et al. 2000, Molinari et al. 2008). 
The sources were placed in Fig.~\ref{menv_lbol} under the assumption that the most luminous object 
in each core (dominating $L_{bol}$ in Table~\ref{ngc3576_tab2}) coincides with the column density peak 
of that core (dominating $M_{env}^{0.03\rm{pc}}$). 
Evolutionary tracks, computed assuming each protostar forms from a bounded condensation of finite initial mass  $M_{env}(0) $
and has $L_{bol} = GM_{\star}(t)\dot{M}_{acc}(t)/R_{\star}(t) + L_{\star}(t)$, where 
$R_{\star}(M_{\star})$ is the protostellar radius and 
$L_{\star}$ is the interior stellar luminosity (Stahler 1988, Hosokawa \& Omukai 2008), 
are shown.  
The mass accretion rate and the envelope mass were assumed to be related by $\dot{M}_{acc} (t) = \epsilon\,M_{env}(t)/\tau$, 
where $\epsilon = 50\% $ is the typical star formation efficiency for individual cores (cf. Matzner \& McKee 2000)
and $\tau = 10^5$~yr is the characteristic timescale of protostellar evolution, leading to $\dot{M}_{acc} (t)$ and 
$M_{env}(t) $ functions declining exponentially with time (see Bontemps et al. 1996). 
It can be seen that the two tracks corresponding to final stellar masses of $15\, M_\odot$ and $50\, M_\odot$, respectively, 
bracket the locations of the NGC~3576 objects in the $M_{env}$--$L_{bol}$ diagram, suggesting that these objects 
will evolve into stars of masses between $\sim 15\, M_\odot $ and $\sim 50\, M_\odot $. 
The relative positions of the NGC~3576 sources in Fig.~\ref{menv_lbol} also support the view that 
S1-M1, S1-M2, S4-M6, and S5-M8 
are observed at significantly earlier evolutionary stages than are S3-M4 and S3-M5, while 
S3-C3 appears to be at an intermediate evolutionary stage. 
Despite relatively large uncertainties, the values derived for both the $M_{env}/L_{bol}^{0.6}$ evolutionary indicator 
(Andr\'e et al. 2000, Bontemps et al. 1996) and the $T_{bol}$ indicator (Myers et al. 1998) 
(see Col.~6 and Col.~7 
of Table~\ref{ngc3576_tab2}, respectively) indeed suggest that the cores S1-M1, S1-M2, S4-M6, S5-M8
contain at least one massive Class~0 object each and that the cores S3-M4, S3-M5  are dominated by a massive 
Class~I object. 
Recall that bona-fide low-mass Class~0 objects are characterized by 
$M_{env}/L_{bol}^{0.6} \simgt 0.4\, M_\odot /L_\odot ^{0.6} $ and $T_{bol} < 70$~K  (cf. Andr\'e et al. 2000). 

Our P-ArT\'eMiS 450~$\mu$m results in NGC~3576 illustrate the unique potential of the future ArT\'eMiS bolometer camera 
for studies of star-forming regions in the southern hemisphere with APEX. 
With a factor of 9 more pixels than P-ArT\'eMiS at 450~$\mu$m, and significantly better detectors, 
ArT\'eMiS will be nearly two orders of magnitude faster than P-ArT\'eMiS for mapping extended fields.
Furhermore, in addition to 450~$\mu$m images, ArT\'eMiS will simultaneously deliver 350~$\mu$m and 200~$\mu$m 
images, which will provide much better sampling of protostellar SEDs near their peak, hence  
improved bolometric luminosities. The colors measured between 450~$\mu$m and 200~$\mu$m 
will also lead to improved dust temperature estimates and will greatly help to discriminate between self-luminous 
accreting protostars and externally heated, starless cores. 
Combining extensive LABOCA surveys at 870~$\mu$m (such as ATLASGAL -- Schuller, Menten et al., in prep.) 
with follow-up imaging at 200--450~$\mu$m with ArT\'eMiS will make it possible to identify and characterize a large number 
of candidate massive protostars in the southern sky.
These candidate protostars can then be studied in greater detail with ALMA in order to shed light on the formation 
process of high-mass stars. In this way, APEX can truly fulfill its role of  ``ALMA pathfinder''.

\begin{acknowledgements}
We are grateful to the APEX Board for awarding us technical time on APEX 
in March 2007. We would like to thank the APEX staff in Chile for their support 
during and prior to the observations. We acknowledge financial support from 
the French national programs on stellar and ISM physics (PNPS and PCMI), as well
as the French National Research Agency (ANR).
\end{acknowledgements}


\normalsize
\begin{thebibliography}{}

\bibitem[Andr{\' e} \& Montmerle(1994)]{Andre94} Andr\'e, P., \& Montmerle, T. 1994, \apj, 420, 837
\bibitem[Andr{\' e}, Ward-Thompson, \& Barsony(2000)Andr{\' e} et al.]{Andre00} Andr{\' e}, P.,
Ward-Thompson,D., \& Barsony, M.\ 2000, in Protostars and Planets IV, Eds V. Mannings et al., 
p. 59
\bibitem[Beltr\'an et al.(2006)]{Beltran06} Beltr\'an, M. T., Brand, J., Cesaroni, R. et al.
2006, \aap, 447, 221
\bibitem[Billot et al.(2006)]{Billot06} Billot, N. et al. 2006, Proc. SPIE, Vol. 6275, p. 62650
\bibitem[Bonnell, Vine, \& Bate(2004)Bonnell et al.]{Bonnell04} Bonnell, 
I.~A., Vine, S.~G., \& Bate, M.~R.\ 2004, \mnras, 349, 735
\bibitem[Bontemps et al.(1996)]{Bontemps96} Bontemps, S., Andr\'e, P., Terebey, S., \& Cabrit, S. 1996, \aap, 311, 858
\bibitem[Caswell(2004)]{Caswell04} Caswell, J.L. 2004, \mnras, 351, 279
\bibitem[De Pree, Nysewander, \& Goss(1999)De Pree et al.]{DePree99} De Pree, C.G., Nysewander, M.C., Goss, W.M. 1999, \aj, 117, 2902
\bibitem[de Wit et al.(2005)]{deWit05} de Wit, W.J., Testi, L., Palla, F. et al. 
2005, \aap, 437, 247
\bibitem[Henning, Michel, \& Stognienko(1995)Henning et al.]{Henning95} Henning, Th., Michel, B., \& Stognienko, R. 1995, 
Planet. Space Sci., 43, 1333
\bibitem[Hidebrand(1983)]{Hildebrand83} Hildebrand, R.H. 1983, \qjras, 24, 267
\bibitem[Hill et al.(2005)]{Hill05} Hill, T., Burton, M., Minier, V. et al. 
2005, \mnras, 363, 405
\bibitem[Hosokawa \& Omukai(2008)]{Hosokawa08} Hosokawa, T., \& Omukai, K. 2008, \apj  ~(astro-ph/arXiv:0806.4122) 
\bibitem[Matzner \& McKee(2000)]{matzner00} Matzner, C. \& McKee, C. 2000, ApJ, 545, 364
\bibitem[Mc Kee \& Tan(2003)]{MT03}Mc Kee, C. \& Tan, J., 2003, ApJ, 585, 850
\bibitem[Molinari et al.(2008)]{Molinari08} Molinari, S., Pezzuto, S., Cesaroni, R. et al. 
2008, \aap, 481, 345  
\bibitem[Myers et al.(1998)]{Myers98}  Myers, P.C., Adams, F.C., Chen, H., \& Schaff, E. 1998, \apj, 492, 703
\bibitem[Peretto et al.(2006)]{Peretto06} Peretto, N., Andr{\' e}, P., \& Belloche, A.\ 2006, \aap, 445, 979 
\bibitem[Peretto et al.(2007)]{Peretto07} Peretto, N., Hennebelle, P., \& Andr{\' e}, P.\ 2007, \aap, 464, 983
\bibitem[Persi et al.(1994)]{Persi94} Persi, P., Roth, M., Tapia, M. et al. 
1994,  \aap, 282, 474
\bibitem[Purcell et al.(2008)]{Purcell08} Purcell, C.R., Minier, V., Longmore, S.N. et al.
2008, \mnras
\bibitem[Rev\'eret et al.(2006)]{Reveret06} Rev\'eret, V., Rodriguez, L.,  Andr\'e, P. et al.
2006,  Proc. SPIE, Vol. 6275, p. 627502
\bibitem[Robitaille et al.(2006)]{Robitaille06} Robitaille, T.P., Whitney, B.A., Indebetouw, R., Wood, K., \& Denzmore, P. 2006, \apjs, 
167, 256
\bibitem[Robitaille et al.(2007)]{Robitaille07} Robitaille, T.P., Whitney, B.A., Indebetouw, R., \& Wood, K. 2007, \apjs, 169, 328
\bibitem[Saraceno et al.(1996)]{Saraceno96} Saraceno, P., Andr\'e, P., Ceccarelli, C. et al. 
1996, \aap, 309, 827
\bibitem[Simoens et al.(2004)]{Simoens04} Simoens, F., Agn\`ese, P. et al. 2004, Proc. SPIE, Vol. 5498, p. 177
\bibitem[Stahler(1988)]{Stahler88} Stahler S.W. 1988, \apj, 332, 804
\bibitem[Talvard et al.(2006)]{Talvard06} Talvard, M., Andr\'e, P., Rodriguez, L., Minier, V.  et al. 2006, Proc. SPIE, Vol. 6275, p. 627503
\bibitem[Zinnecker \& Yorke(2007)]{Zinnecker07} Zinnecker, H., \& Yorke, H.W.  2007, \araa, 45, 481

\end{thebibliography}
\end{document}